# Large Latin American Millimeter Array

Gustavo E. Romero

*Instituto Argentino de Radioastronomía (CONICET, CICPBA), Villa Elisa, Argentina.*

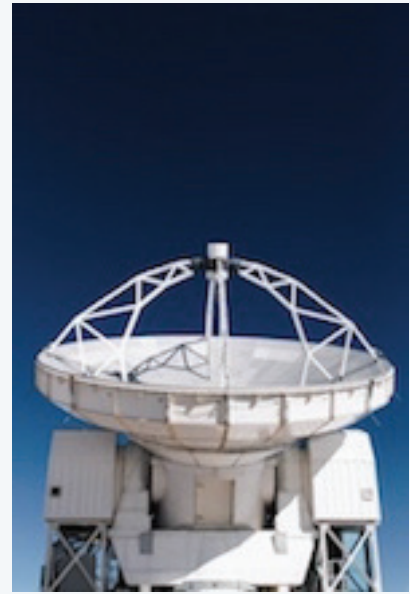

**Abstract**
The Large Latin American Millimeter Array (LLAMA) is a multipurpose single-dish 12 m radiotelescope with VLBI capability under construction in the Puna de Atacama desert in the Province of Salta, Argentina. In this paper I review the project, the instrument, the current status, and the scientific goals of this astronomical collaboration beween Argentina and Brazil.

## Introduction

Astronomy is concerned with the detection and measurement of natural signals emitted from astrophysical sources in the Universe. During most of its long history astronomy was restricted to a very narrow kind of signals: optical radiation. After the work of James Clerk Maxwell in the 1860s, it became clear that light is a phenomenon based on electric and magnetic perturbations and that a continuum of electromagnetic radiation should exist. Between 1886 and 1889 Heinrich Hertz, a young physicist researching at Karlsruhe University, conducted a series of experiments that would prove the existence of electromagnetic waves.

Quite soon radio waves were being used for communications. The discovery in 1902 that these waves are reflected by the ionosphere led physicists to think that astronomical radio signals would not penetrate the atmosphere. Bouncing waves, however, might be used for overseas communication. In the early 1930s, Karl Jansky, an engineer working for Bell Telephone Laboratories, was investigating the static that interfered with short wave transatlantic voice transmissions when serendipitously discovered cosmic radio waves coming from the Galactic center region. He published his results [1] but was not allowed to continue with his astronomical research by the Bell Labs. Grote Reber, an engineer and radio aficionado, followed Jansky's path and built the first orientable radio telescope[1] in the backyard of his house in Wheaton, Illinois. He confirmed Jansky's discovery, found additional astronomical radio sources, and by 1944 published the first radio map of the sky at 160 MHz [2].

Until the end of World War II, Reber remained the world's only radio astronomer. During the late 1940s and early 1950s the field of radio astronomy exploded with the development of new technologies based upon the radio and radar techniques. In 1946 Sir Martin Ryle and D. D. Vonberg made the first astronomical observation using a pair of radio antennas as an interferometer. In 1951 the 21-cm line associated with the emission of hydrogen due to the spin flip of the electron, was detected at Harvard University. Many discoveries followed, including the spiral structure of the Galaxy, extragalactic radio sources, supernova remenants, pulsars, quasars, and more.

In Argentina these discoveries did not go unnoticed. In 1958 a Commission for Radio Astronomy was created at the

---
[1]The instrument had a 9.5-m parabolic dish.

University of Buenos Aires (UBA). Soon things were ready for the establishment of the first radio telescope of South America. The Instituto Argentino de Radioastronomía[2] (IAR) was created in 1962 by Nobel laureate Dr. Bernardo Houssay, then President of the National Research Council of Argentina (CONICET) through an agreement with the Carnegie Institution of Washington. The latter would provide a radio telescope and train engineers that would work on the receiver. The primary goal of the instrument was to survey the HI line in the galactic plane. As first Director of the institute was appointed Carlos Varsavsky, an Argentine physicist who got his PhD at Harvard University and was well-known as the translator to English of Shklovskii's classic book on radio astronomy [3]. The IAR was originally thought as a radio observatory but eventually evolved into a full institute with research staff, students, offices, labs, library, and other facilities (see Figure 1). Till this day it operates two 30-m single-dish radio telescopes, observing at wavelengths of 21 cm in line, timing, or continuum modes.

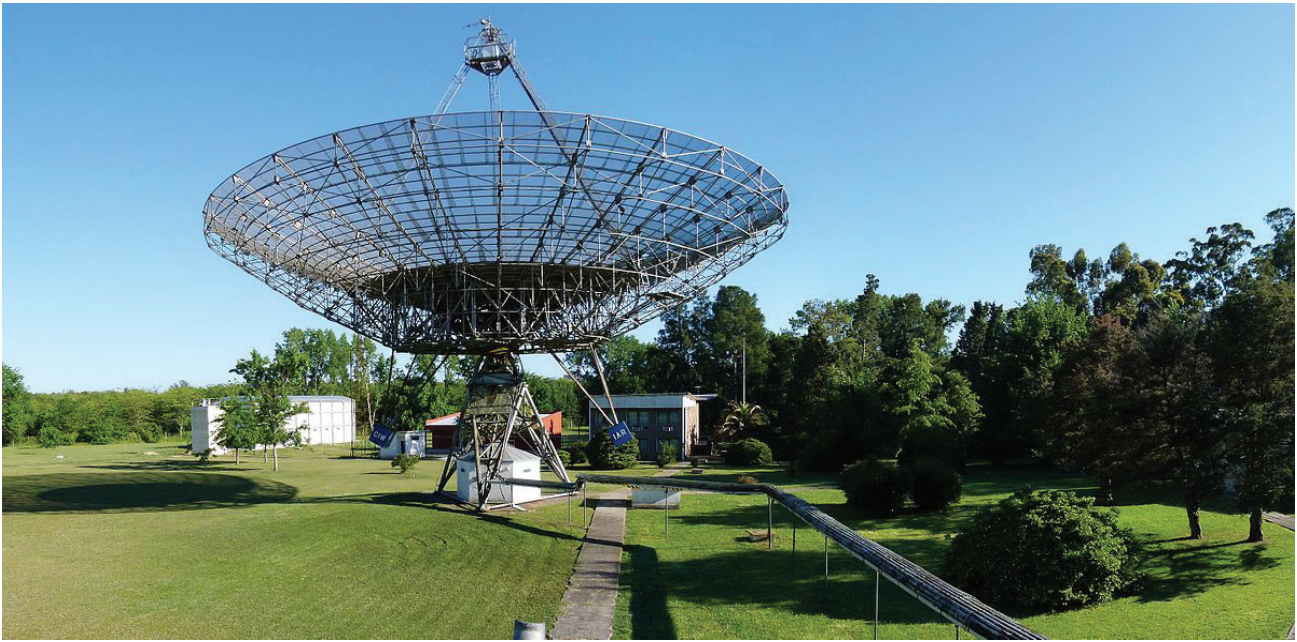

**Figure 1.** View of the Instituto Argentino de Radioastronomía, with one of its two 30-m radio telescopes.

In the late 1960s, while the IAR first telescope commenced to produce results of scientific interest, the Astronomy Committee of the UK's Science Research Council started to analyze the importance of astronomical observations at submillimeter and millimeter wavelengths. In 1975 it was concluded that it would be possible to construct a 15-meter diameter telescope capable of observing at wavelengths down to 750 $\mu$m (Terahertz radiation or THz). The submillimeter radio window (from far infrared to microwaves) is particularly challenging because this radiation is absorbed in the atmosphere by water vapor. At low elevations, where most water vapor resides, the atmosphere is very opaque at submillimeter wavelengths; the abundant water vapor absorbs any incoming submillimeter photons before they can reach the telescope. At higher elevations, however, the water content decreases substantially. By minimizing the atmospheric water vapor, one improves the transparency of the atmosphere and makes astronomical observations possible. It is for this reason that infrared and submillimeter observatories should be built as high as possible and in very dry places: by being above some of the atmosphere, the radiation from astronomical sources is much less attenuated.

There are only a handful sites identified on the Earth that are suitable for submillimeter radio observatories. These include, Mauna Kea (Hawaii, United States), the Atacama Plateau (in the Northwest of Argentina and Northeast of Chile), the South Pole, and Ladakh, India.

The submillimeter telescope recommended by the British Science Research Council would be eventually constructed in Mauna Kea and named the James Clerk Maxwell Telescope. The telescope saw first light in 1987. Since then, many other submillimeter telescopes have been built in the few available sites. The most notorious is the Atacama Large Millimeter Array (ALMA), an astronomical interferometer of 66 radio telescopes in the Atacama Desert (on the Chilean side). At a cost of about USD 1.4 billion, ALMA is the most expensive ground-based telescope in operation in the world. It is a joint enterprise of partners in Europe, the United States, Canada, Japan, South Korea, Taiwan, and Chile. ALMA has a higher sensitivity and higher resolution than any existing submillimeter facility. This is achieved thanks to the large number of antennas and the fact that they can be separated up to 16 km to determine long baselines for interferometry.

At a basic level, interferometry is simply the combining of signals from two different sources. If the two signals are similar then they will combine to make a stronger signal, and if they are not, they will tend to cancel out. This becomes useful for astronomy when two signals are out of synchronicity: then it is possible to shift them (correlate them) so that they will be synchronized. When the signal is strongest, the signals are correctly lined up.

---
[2]Argentine Institiute for Radio Astronomy.

When two radio antennas are aimed in the same direction, they receive the same basic signal, but the signals are out of synchronicity because it takes a bit longer to reach one antenna than the other. That difference depends on the direction of the antennas and their spacing apart from each other. By correlating the two signals, it is possible to determine the location of the signal in the sky very precisely.

Two antennas only give one point in the sky, but dozens of antennas (such as the array at ALMA) can get lots of points, one for each pairing of antennas, creating an image. But this will be a discretely sampled image. If the Earth were fixed in relation to the sky, then a radio image would look like a pointillist painting. Since the Earth rotates with respect to the sky, so as time goes by the relative positions of the antennas shift with respect to an astronomical signal. As the observations are done, the gaps between antennas are filled to create a more solid image. It takes lots of observations and lots of computing power to combine the images in the right way. At ALMA there is a supercomputer that spends all its time correlating signals. The results are superb.

By the time when ALMA started operations in 2011, both Argentina and Brazil had embarked into a joint project to build a submillimeter observatory that might be used to extend ALMA baselines more than 100 km, with the consequent increase in angular resolution. Such project was named Large Latin America Millimeter Array (LLAMA). It is described in the pages that follows.

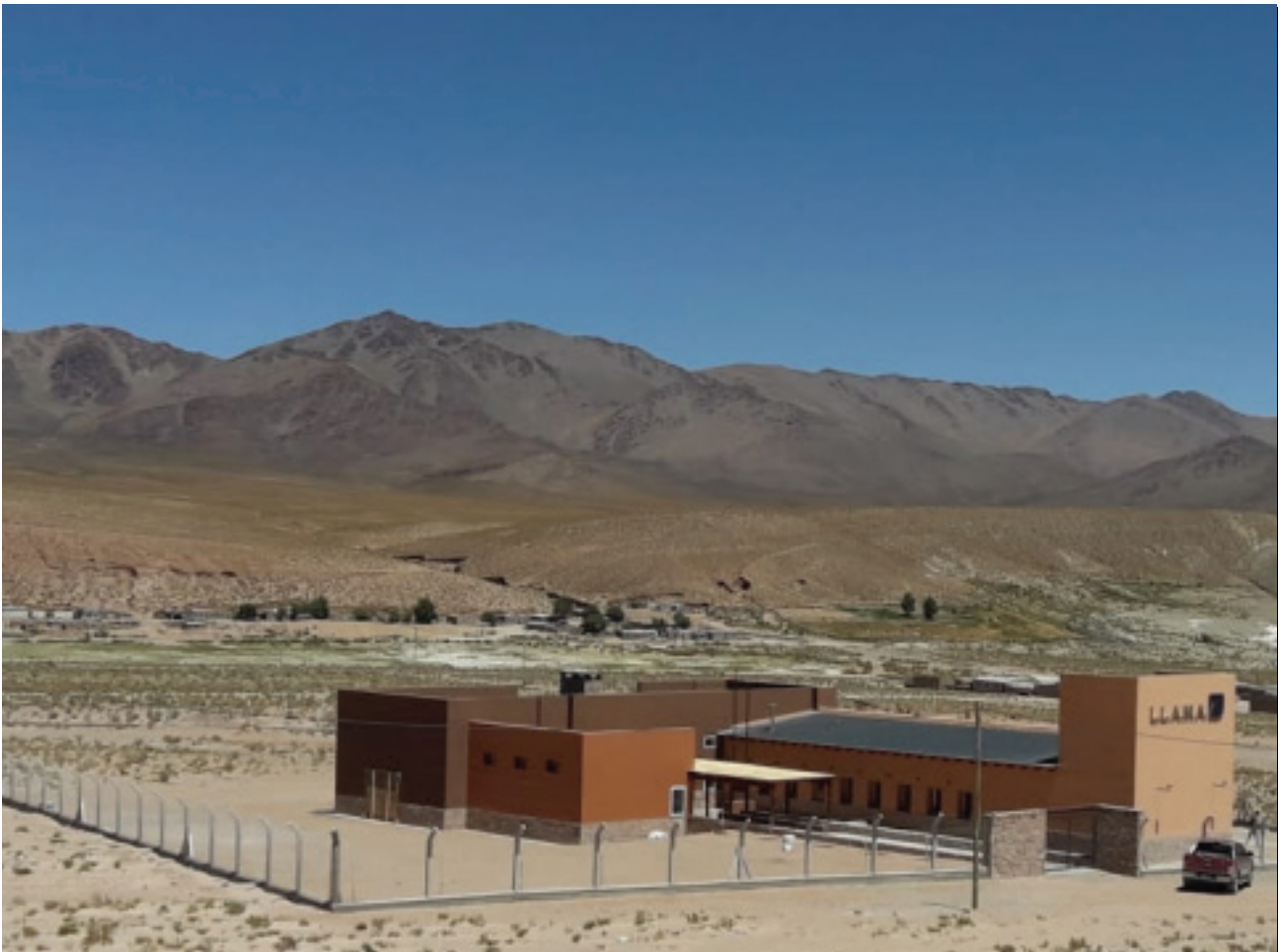

**Figure 2.** LLAMA building for administration, lodging, and labs in San Antonio de los Cobres, Salta. Credit: Gobierno de la Provincia de Salta.

## 1. The project

Campaigns to characterize the potential of the astronomical sites for short wavelengths radio astronomy in the North of Argentina started in the early 2000s. These campaigns were conducted by the IAR and consisted of meteorological (especially measurements of atmospheric opacity) and topographical data collection. In 2007, discussions between scientists from Argentina and Brazil started to shape a joint project for the installation of a submillimeter facility in the Argentine extension of the Atacama desert. Such a facility should be able to operate both in single-dish mode or as a component of a large interferometer array with ALMA or other telescopes. The project was presented to the Argentine scientific community in 2008, during the Annual Meeting of the Argentine Astronomical Society held in San Juan [4]. The formal presentation before the Argentine Science Ministry (MinCyT) was in 2010, while a meeting held at the offices of the Brazilian agency FAPESP, based on São Paulo, in August 2011, marked the formal entrance of Brazil in the project [5]. In 2011 MinCyT ranked LLAMA as its main astronomical project and in 2012 FAPESP approved a grant of about USD 9 M. The final agreement, between MinCyT, FAPESP, and the University of São Paulo (USP), was signed in June 2014.

The site chosen for the telescope was located in the region of Alto Chorrillos, in Salta province. The site is at 4832.5 meters above sea level and about 16 km away from the small town of San Antonio de los Cobres (longitude 66° 28' 29.4" (W), −24° 11' 31.4"(S)). At the end of 2016 construction work on the route to the top of the mountain started. By the end of 2018 the building of the project in San Antonio de los Cobres was completed (see Fig. 2).

Regarding the telescope, it was decided to acquire an antenna and receivers similar to those ones used in ALMA. The antenna was provided by Vertex, the same company that built some of ALMA antennas. It is described in the next section.

## 2. The instrument

Since LLAMA is expected to be a multipurpose instrument that should operate either as a stand-alone telescope or as a part of a larger interferometer, an obvious option was to base the project on the design of an existing and well-tested antenna. An apparent choice was the design of the Atacama Pathfinder Experiment (APEX), a very successful instrument used to test technologies later used in ALMA, see Figures 3 and 4. The APEX telescope is a modified ALMA prototype antenna and is at the site of the ALMA observatory. APEX is designed to work at submillimeter wavelengths, in the 0.2 to 1.5 mm range and its primary goal is to find targets that ALMA will be able to study in greater detail.

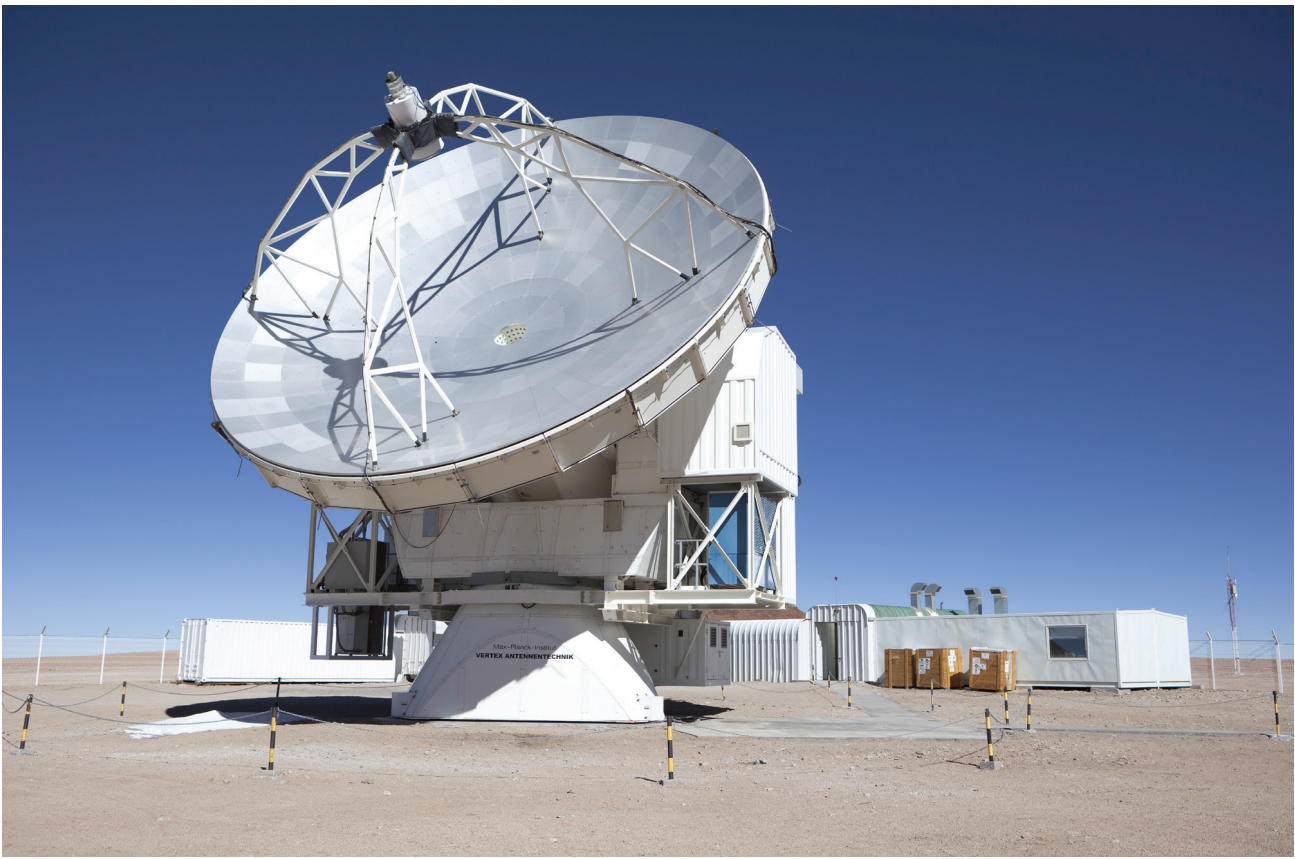

**Figure 3.** The APEX antenna. APEX, the Atacama Pathfinder Experiment, is a collaboration between Max Planck Institut f ur Radioastronomie (MPIfR), Onsala Space Observatory (OSO), and the European Southern Observatory (ESO) aimed at constructing and operating a modified ALMA prototype antenna as a single dish on the high altitude site of Llano Chajnantor in Chile. This instrument essentially operates with the same antenna design that will be used in LLAMA. Credit: ESO.

The antenna for LLAMA was commissioned to the German company Vertex Antennentechnik, GmbH. It is quite similar to APEX. The basic design is shown in Figure 5. The operating frequency range of the antenna will be from 30 GHz to 950 GHz. The antenna has a symmetric paraboloidal reflector with a diameter of 12 m, mounted on an elevation over an azimuth mount. The overall optical layout of the antenna is a Cassegrain configuration with the parameters as shown in Table 1. There are two Nasmyth cabins for single pixel receivers on both sides of the instrument. These heterodyne receivers are listed in Table 2.

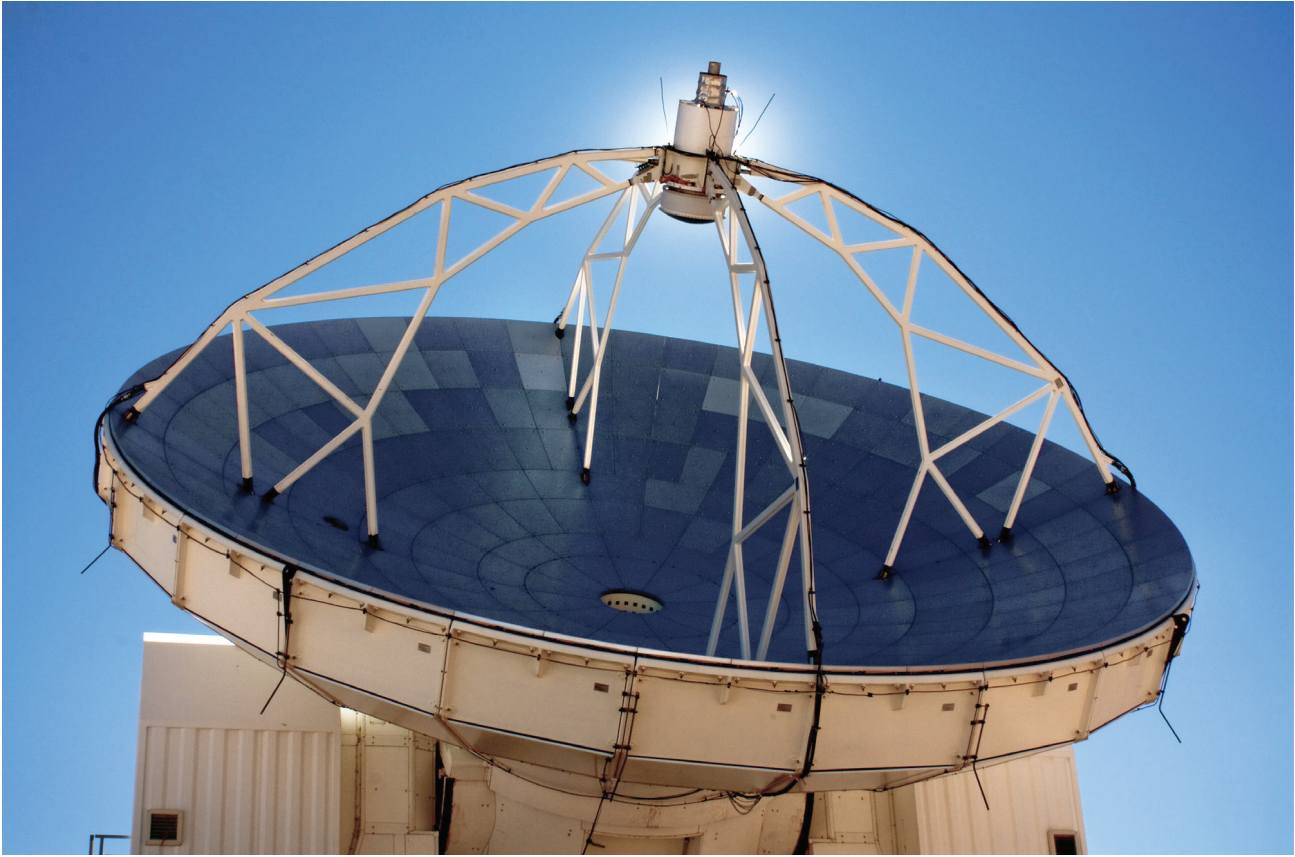

**Figure 4.** Detail of the Vertex antenna. Credit: ESO.

**Table 1.** Optical parameters of LLAMA telescope[6]

| Parameter | Symbol (Name) | Value |
|---|---|---|
| $D$ | Primary Apperture | 12.0 m |
| $f_p$ | Focal Length of Primary | 4.8 m |
| $D_s$ | Secondary Aperture | 0.75 m |
| $M$ | Magnification Factor | 20.0 |
| $\theta_p$ | Primary Angle of Illumination | 128.02° |
| $\theta_s$ | Secondary Angle of Illumination | 7.16° |
| $2c$ | Distance between Primary and Secondary Focus | 6.177 m |
| $H$ | Depth of Primary | 1.875 m |
| $V$ | Primary Vertex Hole Clear Aperture | 0.75 m |
| $W$ | Total Weight | 125 tons |

Major design specifications of the telescope are:

- Capability to operate at an altitude of $\sim 5000$ m.
- Pointing accuracy 2 arcsec (absolute) and 0.6 arcsec (offset pointing).
- Reflector surface accuracy $< 25\ \mu$m.
- Temperature stability of $\pm 1°$ C in receiver cabin.
- Design lifetime $> 20$ years.

The main dish consists of 264 aluminium panels in 8 rings fixed on a carbon fiber reinforced plastic backup structure of 24 sandwich shell segments. Each panel is supported by five vertical (four corners and center) and three horizontal adjustment elements. The panels have been chemically etched to scatter visible and IR radiation and thus allow for daytime observations. To operate at the shorter submillimeter wavelengths, the antenna should have a surface of exceedingly high quality. After a series of high precision adjustments, the surface of the primary mirror will have remarkable precision. Over the 12 m diameter of the antenna, the rms deviation from the perfect parabola is less than 17 thousandths of a millimeter . This is smaller than one fifth of the average thickness of a human hair.

**Table 2.** Single pixel receivers of LLAMA telescope[7]

|  | Name |  |
| --- | --- | --- |
| Band | Frequency (GHz) | Average noise |
| 1 | 35 – 50 | 17 K |
| 3 | 84 – 116 | 35 K |
| 5 | 163 – 211 | 45 K |
| 6 | 211 – 275 | 55 K |
| 7 | 275 – 373 | 70 K |
| 9 | 602 – 720 | 160 K |

The optical system is designed to be as versatile as possible. A maximum of two simultaneous receivers, with dual polarization each, could be used at any given time. Though the optical design of the telescope will not allow Cassegrain/Nasmyth simultaneous observations to be carried out, the system will have the capability of making a fast swap (within a few minutes) between instruments located at the Cassegrain focus (e.g. a MKID camera or an heterodyne array at a given frequency) and those located at a Nasmyth focus.

The different receivers will be contained in cartridges within a cryostat that will cool the electronics to $\sim$ 4 K (see Figure 6). In the Cassegrain focus area there will be a bolometric array. At present, the most promising technology for bolometric cameras employs 300 mK cooled microwave kinetic inductance detectors (MKIDs). These type of extremely cooled arrays of bolometers are similar to those used in the nearby QUBIC telescope [8]. The low frequency receivers are currently being under construction by NOVA[3]. NOVA is also in charge of LLAMA back-end. The cryostat where the cartridges with the receivers are inserted was developed by the National Astronomical Observatory of Japan (NAOJ). It will accommodate up to three ALMA compatible receivers. The cryostat is currently in NOVA for integration. For first light LLAMA is expected to count with receivers operating in Bands 5 and 9 (see Table 2).

In the Cassegrain focus a Water Vapor Radiometer (WVR) could also be located tuned to a frequency of 183 GHz. The WVR would be used to measure the water vapor content of the atmosphere at the time of observation. With this measurement, the observation frequency band can be programmed in real time and the effects of the distortion caused by the atmosphere on the signal phase can be obtained for its later removal, with digital techniques, in the detected signal.

## 3. Science

Much of the electromagnetic radiation in the local Universe arrives in the far-infrared and submillimeter band as thermal radiation from dust. The microwave electromagnetic window in astronomy is particularly suitable for investigations of star forming regions and protostars. Molecular outflows associated with the formation of stars and their winds can be traced by line observations of molecules such as CO, HCN, HCO+, N2H+, CS, and NH3. Interstellar gas clouds are heated by starlight. When they collapse under gravity, they heat up. However, the clouds need to cool in order to form the next generation of stars. The spectral lines that are in the far-infrared and submillimeter bands are the primary coolants for the neutral gas that form stars. Some of the most important cooling lines include H2O, SO2, and CO rotational lines, [CI] [CII], and [NII] fine structure lines. All these lines can be investigated with LLAMA telescope.

ALMA, with its high resolution, has obtained extraordinary images of young stars with protoplanetary disk (see Figure 8). These images reveal the gaps open in the disks by the forming planets. LLAMA can contribute to this kind of studies either observing as a single dish, measuring the rotational velocity of disks in young stellar objects, or as a part of a larger interferometer array helping to increase the angular resolution of the images.

As APEX, LLAMA can be used for the Event Horizon Telescope (ETH), a collaboration that applies submillimeter telescopes to image the supermassive black hole in the galactic center (Figure 7). In the future, this array is also expected to map the black hole in nearby radio galaxies. In 2019 it already has obtained the first image of the shadow of a supermassive black hole at the centered of M87 [9].

LLAMA will also be capable of observing the Sun. Frequencies near the submillimeter range are produced in the lower solar chromosphere or even in the photosphere. LLAMA observations can gather information on the lower solar atmosphere and particle acceleration there. If working in interferometric mode with ALMA or APEX, the array might achieve a spatial resolution of $0.001''$ at wavelengths of about 1 mm. This is tantamount to about 700 m over the solar surface.

---

[3]NOVA stands for 'Nederlandse Onderzoekschool Voor Astronomie', i.e., the 'Netherlands Research School for Astronomy'. It is the alliance of the astronomical institutes of the universities of Amsterdam, Groningen, Leiden, and Nijmegen.

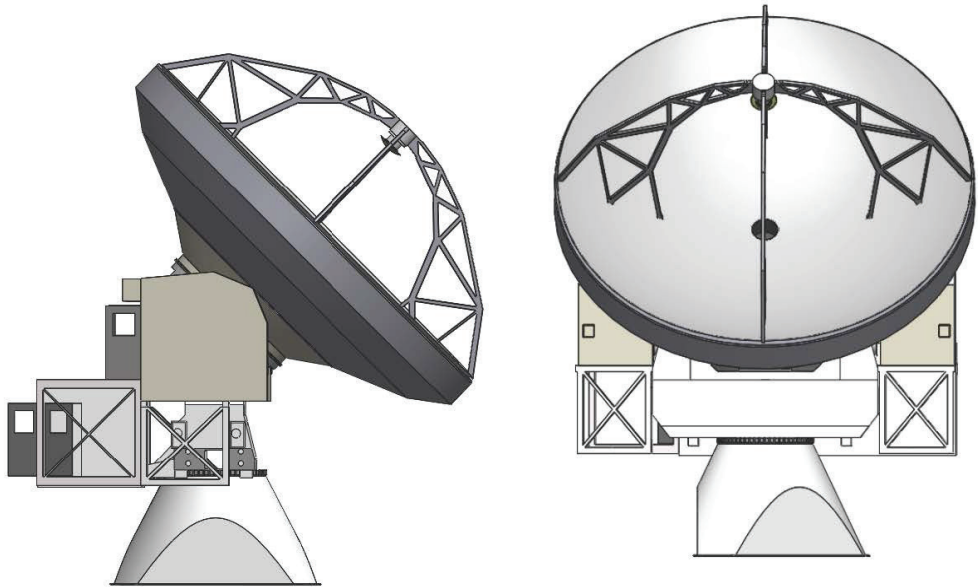

**Figure 5.** Sketch of LLAMA antenna produced by Vertex Antennentechnik, GmbH. Lateral and front views. Credit: Vertex.

Outside the Galaxy, LLAMA can be used to investigate outflows in starbursts, the chemical enrichment of the intergalactic medium around galaxies, dusty Active Galactic Nuclei, and galaxy formation at high redshifts. Dusty star-forming galaxies reprocess stellar light in dust, which radiates in the far-infrared and submillimeter bands. The Milky Way emits about half of its light in this range. The most luminous galaxies in the local Universe, called Ultra-luminous Infrared Galaxies (ULIGS) emit most (up to 99%) of their energy at millimeter and submillimeter wavelengths. These galaxies are natural targets for LLAMA.

Finally, LLAMA also offers a unique opportunity to make research in cosmology. The Cosmic Microwave Background Radiation (CMBR) stems from about 300.000 years after the Big Bang, when the Universe had expanded and cooled to an extent such that atoms could form; leaving light waves to travel freely through space. Until that point, the Universe was ionized, and too dense and hot for light to travel far without interacting with matter. The so-called "last scattering surface" marks the epoch when the Universe became transparent . The radiation from this epoch has been shifted in wavelength because of the continued expansion of the Universe, now appearing in the microwave region of the spectrum, and is observable with millimeter and submillimeter receivers. A major challenge is to detect the polarization of this radiation, in particular the kind of polarization created by gravitational waves traveling through the primordial plasma. This type of polarization is known as the "B mode". Several telescopes, including QUBIC and the Atacama Cosmology Telescope (ACT), are devoted to the detection of this feature in the CMBR. The latter instrument is a six-meter telescope located on Cerro Toco in the Atacama Desert in the north of Chile, near the Llano de Chajnantor Observatory. LLAMA, with its larger surface, might obtain information of CMB fluctuations on smaller scales than these instruments, if an appropriate array of bolometers is located in one of the focus.

## 4. Management

The project is funded by the Secretary of Scientific-Technological Articulation of the Ministry of Science, Technology and Productive Innovation (MINCYT) of Argentina and the Foundation for Research of the State of São Paulo (FAPESP) of Brazil. The construction was supervised by a Steering Committee formed by 6 Argentine and 6 Brazilian scientists, headed by a Director. The first director resigned in late 2017, after the arrival of the telescope to Argentina. The deputy director also resigned in mid 2018, after the instrument was transported to Alto Chorrillos (see Figure 9). The yoke of the antenna was damaged during an accident with one of the trucks. This has produced a delay in the original schedule. Currently the management is being redefined in order to adequately face the final stage of construction and the integration of the telescope. First light is expected in 3-4 years.

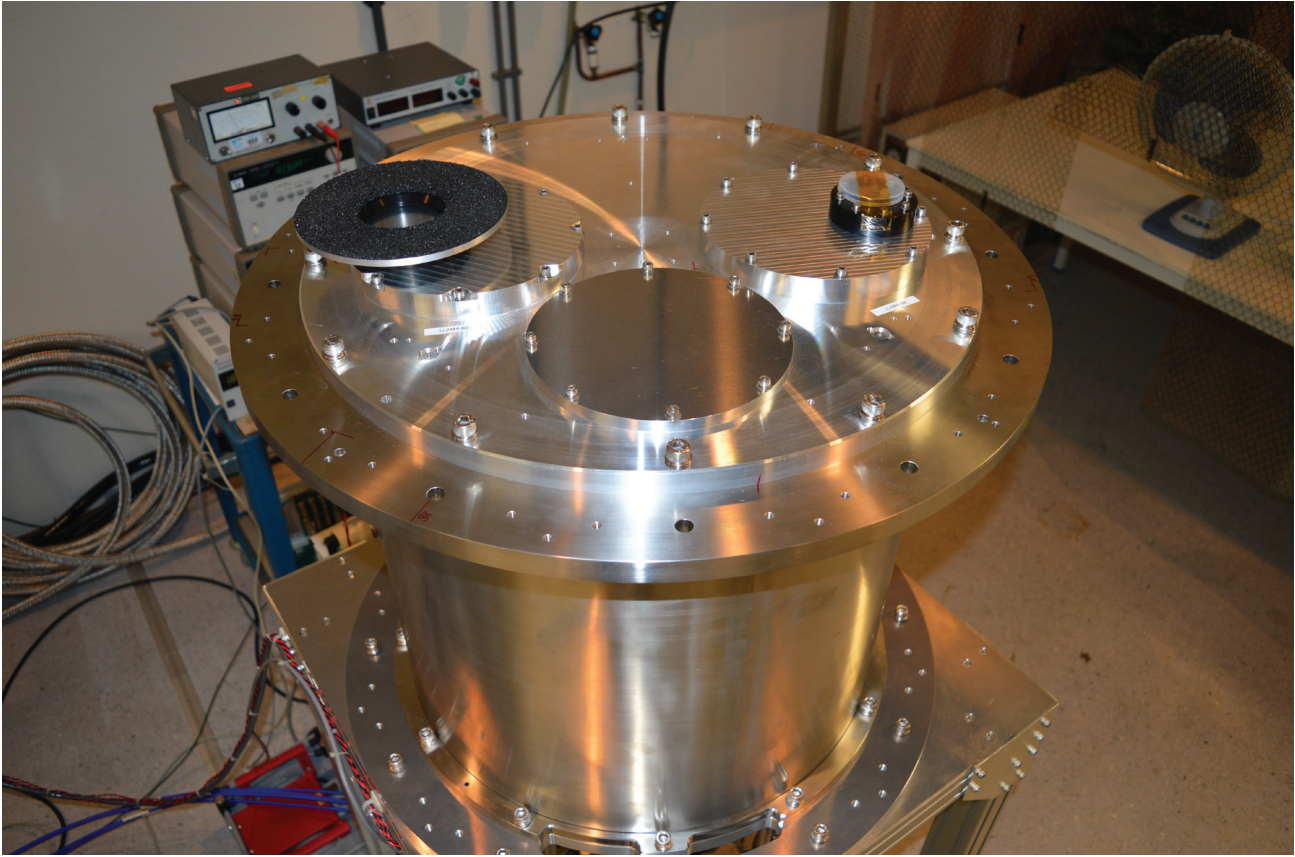

**Figure 6.** Cryostat for the bolometers. Credit: NOVA/ALMA.

## 5. Synergy with other instruments

LLAMA will be part of a larger astronomical center in Salta. In the same summit of the mountain another telescope will be located a few hundred meters away: the Q&U Bolometric Interferometer (QUBIC) [8]. QUBIC is a hybrid telescope of innovative technology for studies of the polarization of the Cosmic Microwave Background (see the article about QUBIC in this same issue). It consists of two arrays of bolometers and an array of horns that allow for observations in two bands: 150 GHz and 220 GHz. The cryostats require similar temperatures to those of LLAMA's receivers (4 and 0.3 K). Both experiments, then, can share supporting infrastructure and personnel.

A number of lower frequency single dish instruments in Argentina can perform observations complementary to those made with LLAMA. ESA's Deep Space Antenna 3 (DSA 3), located in Malargue, Province of Mendoza, is a 35-m antenna that incorporates state-of-the-art technology[10]. Its technical facilities comprise Ka-band reception (31.8 – 32.3 GHz) and X-band transmission and reception. It is prepared to host Ka-band transmission (34.3 – 34.7 GHz) and K-band reception (25.5 – 27 GHz). An agreement between ESA and the Argentine Space Agency (CONAE) grants 10 % of the observing time to Argentina, and part of this time is available to the local astronomical community. Both IAR and the University of Rio Negro are working on a new digital receiver for astronomical purposes. A similar antenna has been built by China in the southern province of Neuquén, and it is also partially available for astronomical uses[11].

Another mm instrument will be located soon in Argentina: the China-Argentina Radio Telescope (CART)[4]. It is a 40-m single dish that will operate at the Felix Aguilar Observatory in San Juan Province. The instrument will be used at low frequencies for VLBI with application to geodesy, but in its final configuration it will be able to operate up to 45 GHz. At such frequencies it will be possible to implement interferometric observations with LLAMA.

A major goal of LLAMA is to form part of a larger array with the ALMA, see Figure 10. In order to operate as a piece in a larger interferometer LLAMA needs an ultra-precise timekeeping. ALMA uses for this purpose an atomic clock powered by a hydrogen maser. A similar device should be implemented in LLAMA to link with other observatories. Also, formal agreements should be signed with ALMA in order to make possible an adequate integration of LLAMA in the array.

---

[4] http://cart.unsj.edu.ar/index.php.

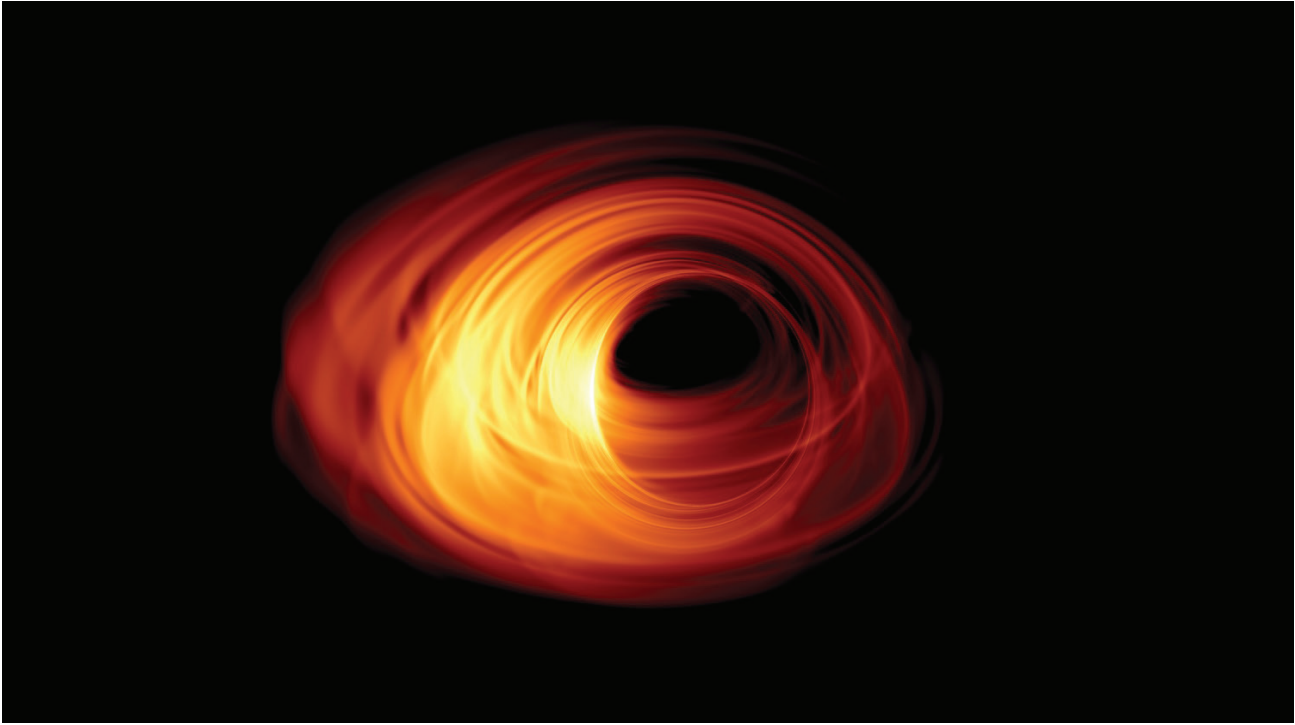

**Figure 7.** Simulated image of an accreting black hole. The event horizon is in the middle of the image, and the shadow can be seen with a rotating accretion disk surrounding it.. Credit: Bronzwaer/Davelaar/Moscibrodzka/Falcke/Radboud University.

## Conclusions

LLAMA telescope is a great challenge for Argentina and Brazil. It represents the most advanced astronomical instrument ever attempted by these countries. Once in operation, it will be a versatile facility that will be used as a single telescope for molecular line and continuum observations and as a component of a larger interferometer for high-resolution imaging of astrophysical sources. This latter utility is by far the most challenging one. It requires to implement technical and scientific methods never tried before in these countries. Advice from and collaboration with ALMA, ESO, NRAO, and other observatories with greater experience in this field will be essential to achieve success. The result, hopefully, will be the formation of a new generation of Latin American radio astronomers and many exciting contributions to our understanding of the Universe.

## Acknowledgments

I thank Dr. Leonardo Pellizza, Dr. Paula Benaglia, and Dr. A. Etchegoyen for support and illuminating discussions about the project and Eng. Juan J. Larrarte for valuable information and help with the multiple tasks associated with the relationship between LLAMA and IAR. I am also grateful to Dr. Paula Benaglia, Eng. Leandro Gracía, Eng. Emiliano Rasztocky, Eng. J.J. Larrarte, Tech. Guillermo Gancio, and Tech. Fernando Hauscarriaga for their contributions to the project. The author was supported by the Argentine agency CONICET (PIP 2014-00338) and the Spanish Ministerio de Economía y Competitividad (MINECO/FEDER, UE) under grant AYA2016-76012-C3-1-P while writing this review.

## References

[1] K. G. Jansky. Radio Waves from Outside the Solar System. *Nature*, 132:66, July 1933.

[2] G. Reber. Cosmic Static. *ApJ*, 100:279, November 1944.

[3] I. S. Shklovskii. *Cosmic radio waves*. Cambridge, Harvard University Press, 1960.

[4] E. M. Arnal, I. F. Mirabel, R. Morras, G. E. Romero, Z. Abraham, E. M. de Gouveira Dal Pino, and J. Lepine. Proyecto "LLAMA". *Boletin de la Asociacion Argentina de Astronomia La Plata Argentina*, 52:357–366, 2009.

[5] E. M. Arnal, R. Morras, G. M. Dubner, E. Giacani, I. F. Mirabel, G. E. Romero, J. R. D. Lepine, Z. Abraham, and E. M. de Gouveia dal Pino. Hacia una integración radioastronómica con Brasil: Proyecto LLAMA (Long Latin American Millimetre Array). *Boletin de la Asociacion Argentina de Astronomia La Plata Argentina*, 54:435–438, 2011.

[6] Vertex Antennentechnik GmbH. *Large Latin-American Millimetric Array Telescope LLAMA – Antenna with Nasmyth Focus – Technical proposal*. Vertex Antennentechnik GmbH, Duisburg, 2014, VA Proj-No.: 21/09087.

[7] LLAMA Collaboration Argentina. *Large Latin America Millimeter Array White Paper*. Preprint (undated).

[8] A. Mennella, P. Ade, G. Amico, D. Auguste, J. Aumont, S. Banfi, G. Barbaràn, P. Battaglia, E. Battistelli, A. Baù, B. Bélier, D. Bennett, L. Bergé, J. Bernard, M. Bersanelli, M. Sazy, N. Bleurvacq, J. Bonaparte, J. Bonis, E. Bunn, D. Burke, D. Buzi, A. Buzzelli, F. Cavaliere, P. Chanial, C. Chapron, R. Charlassier, F. Columbro, G. Coppi, A. Coppolecchia, R. D'Agostino, G. D'Alessandro, P. Bernardis, G. Gasperis, M. Leo, M. Petris, A. Donato, L. Dumoulin, A. Etchegoyen, A. Fasciszewski, C. Franceschet, M. Lerena, B. Garcia, X. Garrido, M. Gaspard, A. Gault, D. Gayer, M. Gervasi, M. Giard, Y. Héraud, M. Berisso, M. González, M. Gradziel, L. Grandsire, E. Guerard, J. Hamilton, D. Harari, V. Haynes, S. Versillé, D. Hoang, N. Holtzer, F. Incardona, E. Jules, J. Kaplan, A. Korotkov, C. Kristukat, L. Lamagna, S. Loucatos, T. Louis, A. Lowitz, V. Lukovic, R. Luterstein, B. Maffei, S. Marnieros, S. Masi, A. Mattei, A. May, M. McCulloch, M. Medina, L. Mele, S. Melhuish, L. Montier, L. Mousset, L. Mundo, J. Murphy, J. Murphy, C. O'Sullivan, E. Olivieri, A. Paiella, F. Pajot, A. Passerini, H. Pastoriza, A. Pelosi, C. Perbost, M. Perciballi, F. Pezzotta, F. Piacentini, M. Piat, L. Piccirillo, G. Pisano, G. Polenta, D. Prêle, R. Puddu, D. Rambaud, P. Ringegni, G. Romero, M. Salatino, A. Schillaci, C. Scóccola, S. Scully, S. Spinelli, G. Stankowiak, M. Stolpovskiy, F. Suarez, A. Tartari, J. Thermeau, P. Timbie, M. Tomasi, S. Torchinsky, M. Tristram, C. Tucker, G. Tucker, S. Vanneste, D. Viganò, N. Vittorio, F. Voisin, R. Watson, F. Wicek, M. Zannoni, and A. Zullo. QUBIC: Exploring the Primordial Universe with the Q&U Bolometric Interferometer. *Universe*, 5:42, January 2019.

[9] Event Horizon Telescope Collaboration, Kazunori Akiyama, Antxon Alberdi, Walter Alef, Keiichi Asada, Rebecca Azulay, Anne-Kathrin Baczko, David Ball, Mislav Balokovíc, John Barrett, Dan Bintley, Lindy Blackburn, Wilfred Boland, Katherine L. Bouman, Geoffrey C. Bower, Michael Bremer, Christiaan D. Brinkerink, Roger Brissenden, Silke Britzen, Avery E. Broderick, Dominique Broguiere, Thomas Bronzwaer, Do-Young Byun, John E. Carlstrom, Andrew Chael, Chi-kwan Chan, Shami Chatterjee, Koushik Chatterjee, Ming-Tang Chen, Yongjun Chen, Ilje Cho, Pierre Christian, John E. Conway, James M. Cordes, Geoffrey B. Crew, Yuzhu Cui, Jordy Davelaar, Mariafelicia De Laurentis, Roger Deane, Jessica Dempsey, Gregory Desvignes, Jason Dexter, Sheperd S. Doeleman, Ralph P. Eatough, Heino Falcke, Vincent L. Fish, Ed Fomalont, Raquel Fraga-Encinas, William T. Freeman, Per Friberg, Christian M. Fromm, José L. Gómez, Peter Galison, Charles F. Gammie, Roberto García, Olivier Gentaz, Boris Georgiev, Ciriaco Goddi, Roman Gold, Minfeng Gu, Mark Gurwell, Kazuhiro Hada, Michael H. Hecht, Ronald Hesper, Luis C. Ho, Paul Ho, Mareki Honma, Chih-Wei L. Huang, Lei Huang, David H. Hughes, Shiro Ikeda, Makoto Inoue, Sara Issaoun, David J. James, Buell T. Jannuzi, Michael Janssen, Britton Jeter, Wu Jiang, Michael D. Johnson, Svetlana Jorstad, Taehyun Jung, Mansour Karami, Ramesh Karuppusamy, Tomohisa Kawashima, Garrett K. Keating, Mark Kettenis, Jae-Young Kim, Junhan Kim, Jongsoo Kim, Motoki Kino, Jun Yi Koay, Patrick M. Koch, Shoko Koyama, Michael Kramer, Carsten Kramer, Thomas P. Krichbaum, Cheng-Yu Kuo, Tod R. Lauer, Sang-Sung Lee, Yan-Rong Li, Zhiyuan Li, Michael Lindqvist, Kuo Liu, Elisabetta Liuzzo, Wen-Ping Lo, Andrei P. Lobanov, Laurent Loinard, Colin Lonsdale, Ru-Sen Lu, Nicholas R. MacDonald, Jirong Mao, Sera Markoff, Daniel P. Marrone, Alan P. Marscher, Iván Martí-Vidal, Satoki Matsushita, Lynn D. Matthews, Lia Medeiros, Karl M. Menten, Yosuke Mizuno, Izumi Mizuno, James M. Moran, Kotaro Moriyama, Monika Moscibrodzka, Cornelia Müller, Hiroshi Nagai, Neil M. Nagar, Masanori Nakamura, Ramesh Narayan, Gopal Narayanan, Iniyan Natarajan, Roberto Neri, Chunchong Ni, Aristeidis Noutsos, Hiroki Okino, Héctor Olivares, Gisela N. Ortiz-León, Tomoaki Oyama, Feryal Özel, Daniel C. M. Palumbo, Nimesh Patel, Ue-Li Pen, Dominic W. Pesce, Vincent Piétu, Richard Plambeck, Aleksandar PopStefanija, Oliver Porth, Ben Prather, Jorge A. Preciado-López, Dimitrios Psaltis, Hung-Yi Pu, Venkatessh Ramakrishnan, Ramprasad Rao, Mark G. Rawlings, Alexander W. Raymond, Luciano Rezzolla, Bart Ripperda, Freek Roelofs, Alan Rogers, Eduardo Ros, Mel Rose, Arash Roshanineshat, Helge Rottmann, Alan L. Roy, Chet Ruszczyk, Benjamin R. Ryan, Kazi L. J. Rygl, Salvador Sánchez, David Sánchez-Arguelles, Mahito Sasada, Tuomas Savolainen, F. Peter Schloerb, Karl-Friedrich Schuster, Lijing Shao, Zhiqiang Shen, Des Small, Bong Won Sohn, Jason SooHoo, Fumie Tazaki, Paul Tiede, Remo P. J. Tilanus, Michael Titus, Kenji Toma, Pablo Torne, Tyler Trent, Sascha Trippe, Shuichiro Tsuda, Ilse van Bemmel, Huib Jan van Langevelde, Daniel R. van Rossum, Jan Wagner, John Wardle, Jonathan Weintroub, Norbert Wex, Robert Wharton, Maciek Wielgus, George N. Wong, Qingwen Wu, Ken Young, André Young, Ziri Younsi, Feng Yuan, Ye-Fei Yuan, J. Anton Zensus, Guangyao Zhao, Shan-Shan Zhao, Ziyan Zhu, Juan-Carlos Algaba, Alexander Allardi, Rodrigo Amestica, Jadyn Anczarski, Uwe Bach, Frederick K. Baganoff, Christopher Beaudoin, Bradford A. Benson, Ryan Berthold, Jay M. Blanchard, Ray Blundell, Sandra Bustamente, Roger Cappallo, Edgar Castillo-Domínguez, Chih-Cheng Chang, Shu-Hao Chang, Song-Chu Chang, Chung-Chen Chen, Ryan Chilson, Tim C. Chuter, Rodrigo Córdova Rosado, Iain M. Coulson, Thomas M. Crawford, Joseph Crowley, John David, Mark Derome, Matthew Dexter, Sven Dornbusch, Kevin A. Dudevoir, Sergio A. Dzib, Andreas Eckart, Chris Eckert, Neal R. Erickson, Wendeline B. Everett, Aaron Faber, Joseph R. Farah, Vernon Fath, Thomas W. Folkers, David C. Forbes, Robert

Freund, Arturo I. Gómez-Ruiz, David M. Gale, Feng Gao, Gertie Geertsema, David A. Graham, Christopher H. Greer, Ronald Grosslein, Frédéric Gueth, Daryl Haggard, Nils W. Halverson, Chih-Chiang Han, Kuo-Chang Han, Jinchi Hao, Yutaka Hasegawa, Jason W. Henning, Antonio Hernández-Gómez, Rubén Herrero-Illana, Stefan Heyminck, Akihiko Hirota, James Hoge, Yau-De Huang, C. M. Violette Impellizzeri, Homin Jiang, Atish Kamble, Ryan Keisler, Kimihiro Kimura, Yusuke Kono, Derek Kubo, John Kuroda, Richard Lacasse, Robert A. Laing, Erik M. Leitch, Chao-Te Li, Lupin C. C. Lin, Ching-Tang Liu, Kuan-Yu Liu, Li-Ming Lu, Ralph G. Marson, Pierre L. Martin-Cocher, Kyle D. Massingill, Callie Matulonis, Martin P. McColl, Stephen R. McWhirter, Hugo Messias, Zheng Meyer-Zhao, Daniel Michalik, Alfredo Montaña, William Montgomerie, Matias Mora-Klein, Dirk Muders, Andrew Nadolski, Santiago Navarro, Joseph Neilsen, Chi H. Nguyen, Hiroaki Nishioka, Timothy Norton, Michael A. Nowak, George Nystrom, Hideo Ogawa, Peter Oshiro, Tomoaki Oyama, Harriet Parsons, Scott N. Paine, Juan Peñalver, Neil M. Phillips, Michael Poirier, Nicolas Pradel, Rurik A. Primiani, Philippe A. Raffin, Alexandra S. Rahlin, George Reiland, Christopher Risacher, Ignacio Ruiz, Alejandro F. Sáez-Madaín, Remi Sassella, Pim Schellart, Paul Shaw, Kevin M. Silva, Hotaka Shiokawa, David R. Smith, William Snow, Kamal Souccar, Don Sousa, T. K. Sridharan, Ranjani Srinivasan, William Stahm, Anthony A. Stark, Kyle Story, Sjoerd T. Timmer, Laura Vertatschitsch, Craig Walther, Ta-Shun Wei, Nathan Whitehorn, Alan R. Whitney, David P. Woody, Jan G. A. Wouterloot, Melvin Wright, Paul Yamaguchi, Chen-Yu Yu, Milagros Zeballos, Shuo Zhang, and Lucy Ziurys. First M87 Event Horizon Telescope Results. I. The Shadow of the Supermassive Black Hole. *The Astrophysical Journal*, 875(1):L1, Apr 2019.

[10] P. Benaglia, N. Casco, S. Cichowolski, A. Cillis, B. García, D. Ravignani, E. M. Reynoso, and G. de la Vega. The antenna DSA 3 and its potential use for Radio Astronomy. *Boletin de la Asociacion Argentina de Astronomia La Plata Argentina*, 54:447–450, 2011.

[11] M. Colazo. Las antenas de espacio profundo en la Argentina. *Boletin de la Asociacion Argentina de Astronomia La Plata Argentina*, 60:65–66, August 2018.

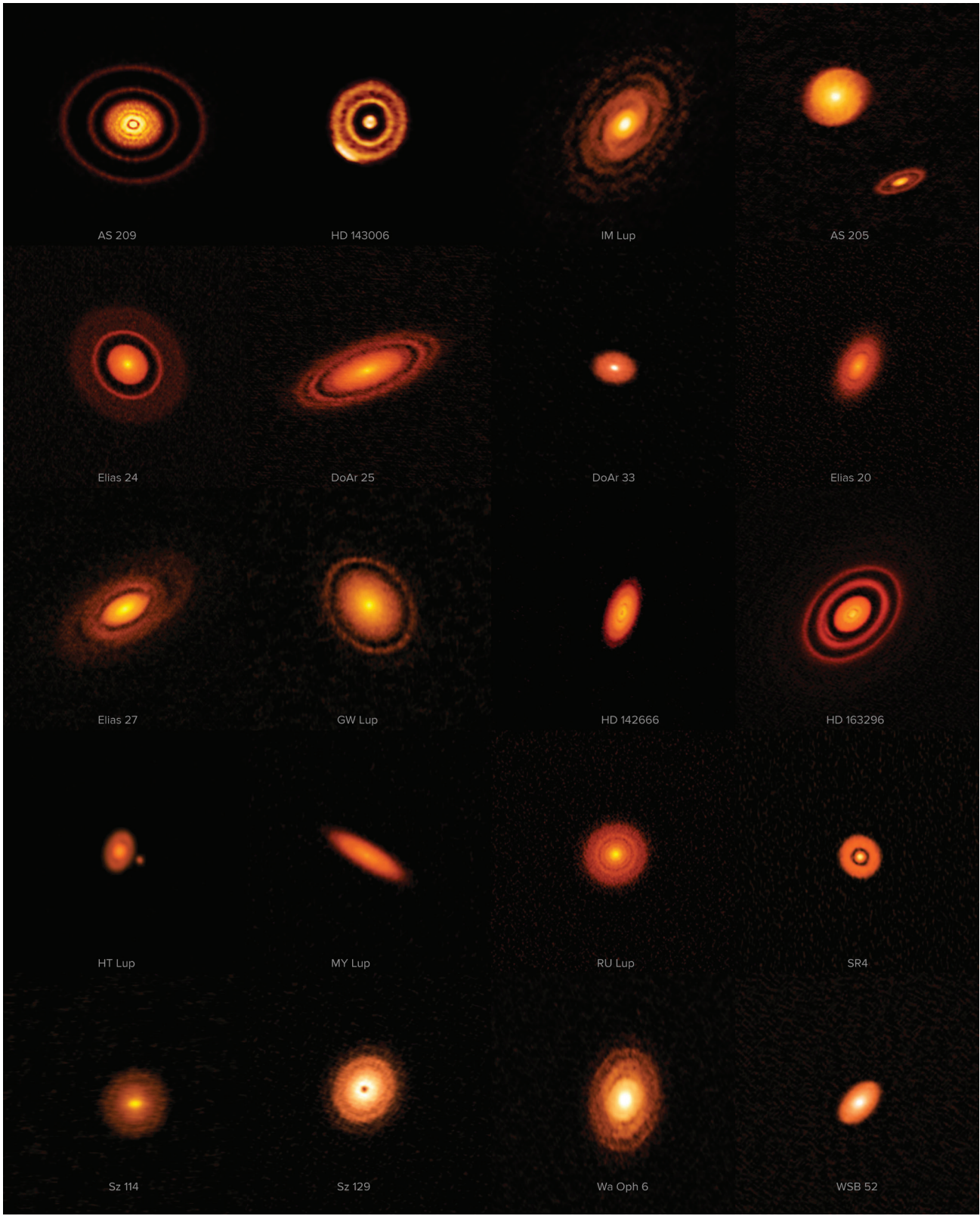

**Figure 8.** Protoplanetary discs around young stars captured by ALMA's first Large Program, known as the Disk Substructures at High Angular Resolution Project (DSHARP). Credit: ALMA (ESO/NAOJ/NRAO), S. Andrews et al.; NRAO/AUI/NSF, S. Dagnello .

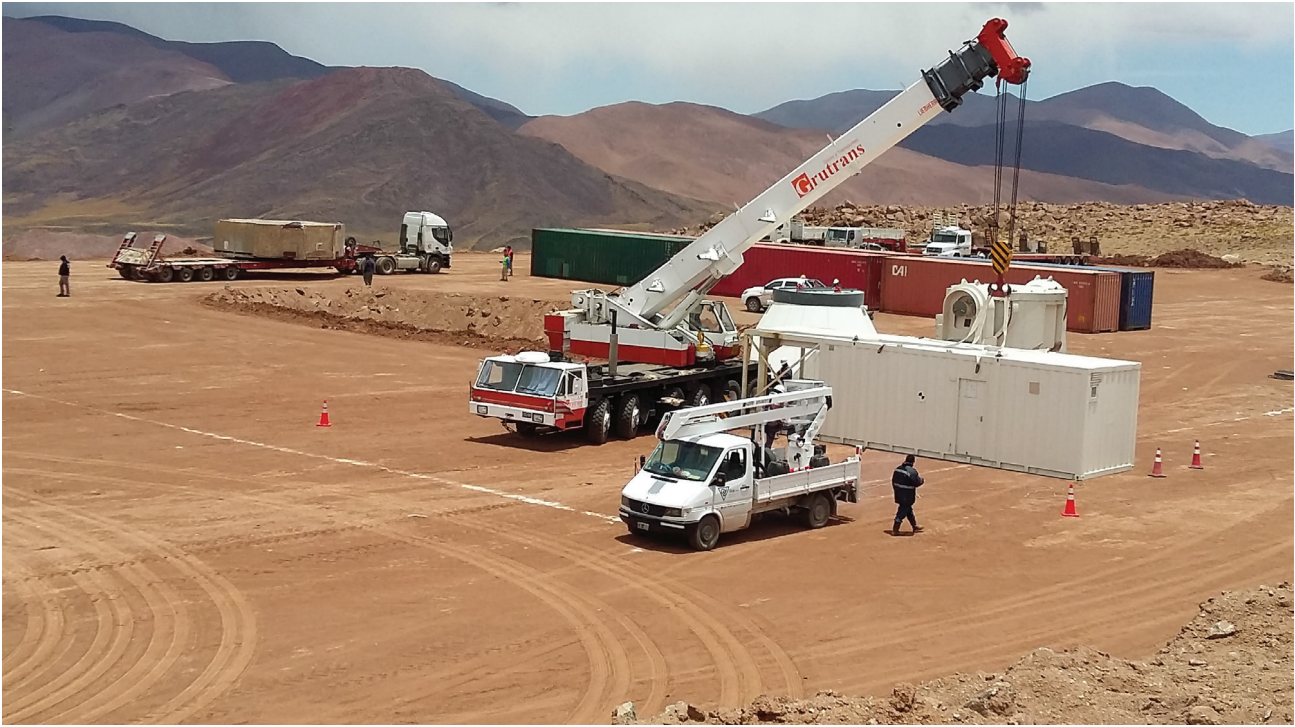

**Figure 9.** View of the installation of different components of LLAMA antenna in Alto Chorrillos . Credit: Fundación CAPACIT-AR and Instituto Geonorte, INENCO (CONICET).

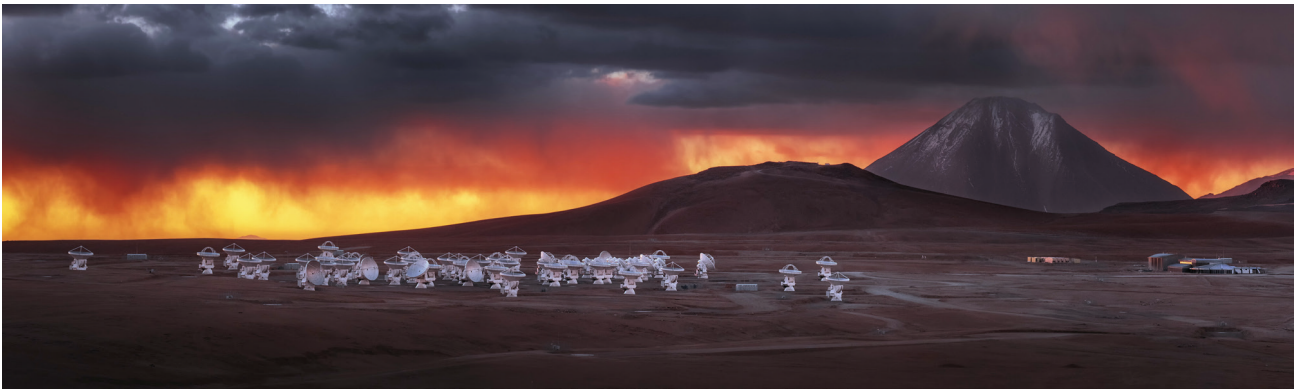

**Figure 10.** View of the Atacama Large Millimeter Array (ALMA). Credit: Y. Beletsky/ESO

## Bio

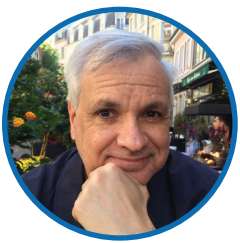

### Gustavo E. Romero

Gustavo E. Romero is a Full Professor of Relativistic Astrophysics at the University of La Plata and Superior Researcher of the National Research Council of Argentina (CONICET). He is a former President of the Argentine Astronomical Society and is the current Director of the Argentine Institute for Radio Astronomy (IAR). Romero is one of the most frequently cited scientists of Argentina. He has published more than 450 papers on astrophysics, gravitation, the foundations of physics, philosophy, and 12 books. Most of his research focuses on black hole physics, gamma-ray astrophysics, gravitation, cosmic rays, and ontological problems of spacetime theories. He has received numerous awards in recognition of his work, including the Houssay Prize (twice), the Gaviola Award from the National Academy of Sciences of Argentina, and the Helmholtz International Award. Among his books we can mention *Introducción a la Astrofísica Relativista* (Universitat de Barcelona Press, 2011 in collaboration with J.M. Paredes), *Introduction to Black Hole Astrophysics* (Springer, 2013 in collaboration with G. Vila) and *Scientific Philosophy* (Springer, 2018).